# Synthesis, Functionalization and Properties of Uniform Europium-doped Sodium Lanthanum Tungstate and Molybdate (NaLa(XO$_4$)$_2$, X= Mo,W) probes for Luminescent and X-ray Computed Tomography Bioimaging

**Mariano Laguna,[a] Nuria O. Nuñez,[a] Ana I. Becerro,[a] Gabriel Lozano,[a] Maria Moros,[b] Jesús M de la Fuente,[b] Ariadna Corral,[c] Marcin Balcerzyk,[c] and Manuel Ocaña[a,*]**

[a]Instituto de Ciencia de Materiales de Sevilla, CSIC-US, Américo Vespucio 49, 41092, Isla de la Cartuja, Sevilla, Spain

[b]Instituto de Ciencia de Materiales de Aragón, CSIC/UniZar and CIBER-BBN, Edificio I+D, Mariano Esquillor s/n, 50018 Zaragoza, Spain

[c]Centro Nacional de Aceleradores (CNA) (Universidad de Sevilla, Junta de Andalucía, CSIC), c/ Thomas Alva Edison 7, 41092, Isla de la Cartuja, Sevilla, Spain

[*] corresponding author:

Manuel Ocaña

e-mail:mjurado@icmse.csic.es

Phone number: +34 954489533

**Abstract:**

A one-pot simple procedure for the synthesis of uniform, ellipsoidal $Eu^{3+}$-doped sodium lanthanum tungstate and molybdate ($NaLa(XO_4)_2$, X = W, Mo) nanophosphors, functionalized with carboxylate groups, is described. The method is based on a homogeneous precipitation process at 120 ºC from appropriate $Na^+$, $Ln^{3+}$ and tungstate or molybdate precursors dissolved in ethylene glycol/water mixtures containing polyacrylic acid. A comparative study of the luminescent properties of both luminescent materials as a function of the $Eu^{3+}$ doping level has been performed to find the optimum nanophosphor, whose efficiency as X-ray computed tomography contrast agent is also evaluated and compared with that of a commercial probe. Finally, the cell viability and colloidal stability in physiological pH medium of the optimum samples have also been studied to assess their suitability for biomedical applications.

# 1. Introduction

The interest in Ln-based phosphors (Ln = lanthanide), usually consisting of nanoparticles made up of a crystalline host (oxides, fluorides, phosphates and vanadates [1]) doped with active $Ln^{3+}$ cations, has greatly increased during the last years because of their applicability as luminescent probes for several biomedical applications among which, imaging is the one most commonly investigated [2]. Such interest mainly arises from the higher chemical and optical stability of these phosphors when compared with other available luminescent materials along with their lower toxicity [3]. Other advantages of Ln-based nanoparticles include high luminescence quantum yield and narrow emission bands, which result in a high selectivity for bioassays [3].

The luminescent characteristics of this kind of materials are determined by the nature of the doping cations. Thus, down-shifting (emissions at lower energy than that of excitation) or up-converting (emissions at higher energy than that of excitation) phosphors can be obtained by a proper selection of the emitting cations. In the case of down-shifting phosphors, the most appropriated Ln doping cation for bioapplications is red emitting $Eu^{3+}$, since possible interferences caused by the autofluorescence of tissues are minimized in this wavelength range [4]. Although $Eu^{3+}$-based phosphors are not useful for *in vivo* applications because of the need of UV excitation, they have been used for many *ex vivo* and *in vitro* biomedical applications [5].

It is important to mention that for this kind of uses, the bioprobes must fulfil several requirements related to: *i)* uniformity in size and shape, since these factors affect the interactions with cell and tissues [6], *ii)* particle size, which must be kept within the nanometer range, especially for *in vivo* applications, to control the circulation behaviour of the nanoparticles in the body, their biodistribution and their excretion pathway [6], *iii)* toxicity, since the nanoprobes must be obviously non-toxic and *iv)* colloidal stability, which must be high in physiological media [7]. To meet the latter criteria, the nanoparticles may require surface coating with organic molecules having different functional groups (carboxylate, amino, imino, etc.), which confer to them a high colloidal stability by increasing their surface charge or by steric hindrance [8]. In addition, a high emission intensity is required for luminescent imaging. However, the absorption coefficient of the $Ln^{3+}$ cations is low, which results in low intensity of the emissions [9]. To overcome this problem, several strategies can be adopted. One of them takes advantage of the large absorption coefficient of some inorganic anions, which are able to transfer the absorbed energy to the active Ln ions thus increasing the intensity of luminescence [9]. Among these anions are tungstates and molybdates, which efficiently absorb UV light that can be transferred to some $Ln^{3+}$ ions such as $Eu^{3+}$ [10-13]. In particular, the scheelite family having general formula $MR(XO_4)_2$ (M = $Na^+$ or $K^+$, R = rare earth cation and X = W or Mo) has been shown to be a promising host for doping with $Ln^{3+}$ cations resulting in high efficient phosphors [14]. It should be noted that the composition of these phosphors confer to them an additional and useful functionality as contrast agent for X-ray computed tomography (CT), a commonly used technique for biomedical imaging [1]. The reason for that is that they contain elements with high atomic number (lanthanides, molybdenum and mainly tungsten), which, consequently, provide them with high X-ray attenuation capacity [1].

In spite of such advantages, the biomedical applications of $MR(XO_4)_2$-based compounds has been rarely addressed. Only a few reports have been found in literature for some members of the family like $NaLa(MoO_4)_2$ [15], $NaGd(MoO_4)_2$ [16] and $NaGd(WO_4)_2$ [17,18], whereas no reports have been found for $NaLa(WO_4)_2$. A possible reason for that might be the unavailability of synthesis methods to produce this material with the needed characteristics mentioned above. Most previous reported methods for the synthesis of uniform NaLa(WO$_4$)$_2$ particles resulted in micron sized entities, [19-25] which do not meet the size criteria. Only Wang et al., recently reported a procedure, which produced Dy and Sm doped NaLa(WO$_4$)$_2$ nanocubes based on a solvothermal reaction using oleic acid as dispersing and capping agent [26]. However, the resulting particles were covered by oleic acid molecules making them hydrophobic and therefore, not useful for bioapplications, and they have to be further treated following several procedures (for example, a 5-min treatment with NOBF$_4$) to remove the oleic acid cover [27]. Therefore, the development of synthesis procedures yielding uniform NaLa(WO$_4$)$_2$ nanoparticles doped with luminescent Ln$^{3+}$ cations having colloidal stability in physiological media is highly desirable.

In this work, we describe for the first time a one-pot simple procedure for the synthesis of uniform Eu$^{3+}$-doped NaLa(WO$_4$)$_2$ nanophosphors with ellipsoidal shape and functionalized with carboxylate groups provided by polyacrylic acid (PAA). This polymer has been suggested as one of the most useful functional coatings for biomedical applications due to its high affinity to the cellular membranes [8]. The employed method is based on a homogeneous precipitation process at 120ºC from appropriate Na$^+$, Ln$^{3+}$ and tungstate precursors dissolved in ethylene glycol (EG)/water mixtures containing PAA molecules. We also show that this method can be easily adapted for the synthesis of isomorphous Eu$^{3+}$-doped NaLa(MoO$_4$)$_2$ nanoparticles. A comparative study of the luminescent properties of both luminescent materials as a function of the Eu$^{3+}$ doping level has also been conducted to find the optimum nanophosphor, whose efficiency as CT contrast agent is evaluated and compared with that of a commercial probe. Finally, the cell viability and colloidal stability in a physiological pH media of the optimum samples have also been studied to assess their suitability for biomedical applications.

## 2. Materials and Methods

*2.1. Materials*

Lanthanum (III) chloride hydrate (LaCl$_3$·xH$_2$O, Aldrich, 99%) and europium (III) chloride hexahydrate (EuCl$_3$·6H$_2$O, Aldrich 99%) were selected as Ln precursors whereas sodium tungstate (Na$_2$WO$_4$, Aldrich, ≥ 98%) and sodium molybdate (Na$_2$MoO$_4$, Aldrich, ≥ 98%) were used as tungstate and molybdate source, respectively, both being also the Na$^+$ source. Polyacrylic acid polymer (PAA, average Mw~1800, Aldrich) was used for functionalization, whereas ethylene glycol (EG, Aldrich, ≥ 99.5%) or EG/water (Milli-Q) mixtures were used as solvents. The colloidal stability study was performed using 2-morpholinoethanesulfonic acid (MES, Aldrich, 99%). The commercial CT contrast agent used for comparison purposed was Iohexol (Sigma Aldrich, analytical standard, ≥ 95%).

*2.2. Synthesis and functionalization of NaLa(WO$_4$)$_2$ nanoparticles*

NaLa(WO$_4$)$_2$ nanoparticles were prepared according to the following procedure. First, LaCl$_3$ was dissolved in ethylene glycol (2.5 cm$^3$) under magnetic stirring while heating the vial at ~80ºC to facilitate the dissolution process. The resulting solution (A) was left to cool down to room temperature. In a separate vial, Na$_2$MoO$_4$ was dissolved in 2.5 cm$^3$ of EG or different EG/H$_2$O mixtures (solutions B). In some experiments, different amounts of PAA were also added to the latter solution. Solutions A and B were then mixed together under magnetic stirring and the final mixtures (total volume = 5 cm$^3$) were subsequently aged for 20 hours in tightly closed test tubes using an oven preheated at 120°C. The resulting dispersions were cooled down to room temperature, centrifuged to remove the supernatants and washed, twice with ethanol and once with distilled water. Finally, the precipitates were redispersed in Milli-Q water or, for some analyses, dried at room temperature.

*2.3. Synthesis and functionalization of NaLa(MoO$_4$)$_2$ nanoparticles*

The same procedure described above for NaLa(WO$_4$)$_2$ was used for the synthesis of NaLa(MoO$_4$)$_2$ nanoparticles slightly modified by substituting the tungstate precursor by Na$_2$MoO$_4$.

*2.4. Synthesis and functionalization of Eu$^{3+}$-doped NaLa(XO$_4$)$_2$ nanoparticles (X = W or Mo)*

To obtain Eu$^{3+}$-doped NaLa(XO$_4$)$_2$ nanoparticles (X = W or Mo) functionalized with PAA (NaLa(XO$_4$)$_2$@PAA), we proceeded as described above for the case of the undoped systems but incorporating the desired amount of the doping cations to the starting LaCl$_3$ solution. The total cations concentration in these solutions was kept constant (0.03 mol dm$^{-3}$). The Eu/(Eu+La) molar ratio was varied from 3 to 27% in order to investigate the effects of the doping level on the luminescent properties of the precipitated nanoparticles.

*2.5. Characterization*

The shape of the nanoparticles was examined by transmission electron microscopy (TEM, Philips 200CM) by depositing a droplet of an aqueous suspension of the samples on a copper grid coated with a transparent polymer. Particle size was measured on several hundreds of particles from TEM micrographs.

The crystalline structure of the prepared particles was identified by X-ray diffraction (XRD, Panalytical X´Pert Pro with an X-Celerator detector). Unit cell parameters were determined from the XRD data (collected at intervals of 0.02° 2θ for an accumulation time of 1000 s) by Rietveld refinement using the X´Pert High Score Plus software. The crystallite size was estimated from the most intense XRD peak (200) located at 2θ ~ 33º in the pattern of the NaLa(XO$_4$)$_2$ samples (X = W or Mo) by using the Scherrer formula.

The infrared spectra (FTIR) of the nanophosphors diluted in KBr pellets were recorded in a JASCO FT/IR-6200 Fourier Transform spectrometer. Thermogravimetric analyses (TGA) were performed in air at a heating rate of 10°C min$^{-1}$, using a Q600 TA instrument.

For a proper comparison of the luminescent properties of the nanophosphors containing different Eu$^{3+}$ doping levels, aqueous suspensions of the nanoparticles having the same concentration (1 mmol dm$^{-3}$) were prepared. The excitation and emission spectra of these dispersed samples were measured in a Horiba Jobin Yvon spectrofluorimeter (Fluorolog FL3-11) operating in the front face mode. Excitation and emission slits of 1 nm and 2 nm were used, respectively, for the excitation spectra and vice versa for the emission ones. Measurements of lifetimes associated to the transitions of the Eu$^{3+}$ cation were obtained from luminescence decay curves recorded in the same equipment using a pulsed lamp. The decay curves were registered for the most intense Eu$^{3+}$ emission band at 610 nm. Absolute values of photoluminescence quantum yield were measured from the ratio between light emitted and light absorbed by the different nanophosphor powders using an integrating sphere in a fluorimeter (Edinburgh FLS1000). A powder tray (1 cm diameter) made from polytetrafluoroethylene was used as sample holder. Powder tray was filled with the similar amount of material and covered with a quartz lid to avoid spillage into the sphere. A plug made from the same material that is used for the sphere, also covered with a quartz lid, was used as a blank. Spectra were recorded using an excitation wavelength that corresponds to the maximum of the energy transfer band, which is 262 nm for the tungstate-based nanophosphor and 290 nm for molybdate one. Thus, measurements comprise processes from the absorption of UV light by the tungstate/molybdate matrix, the transfer to Eu$^{3+}$, and the subsequent outcoupling of the emitted light. Typical experimental error in these measurements is 1%.

Photographs showing the luminescence of aqueous suspensions of the Eu$^{3+}$-doped nanophosphors were taken under illumination with an ultraviolet lamp ($\lambda$ = 254 nm).

For the evaluation of the CT contrast efficiency, aqueous dispersions containing different concentrations of the optimized nanophosphors or Iohexol, in 1.5 mL Eppendorf tubes were prepared. The dispersions were placed in a vortex device for 2 min prior to analyses. Each sample (200 µL aliquot) was placed in a multiwell microplate, along with a Milli-Q water sample for calibration. CT images were acquired with a NanoSPECT/CT® (Bioscan). Acquisition parameters included a 106 mA current for a voltage of 65 kV, exposure time per projection of 1500 ms and 360 projections per rotation. The image length was 6 cm, with a pitch of 1. The total acquisition time required was 18 min. The image was reconstructed with Vivoquant image processing software (Invicro), with the exact cone beam filtered back projection algorithm and the Shepp Logan 98 % filter. The resulting image pixel size was uniform in three dimensions at 0.2 mm. Images were analyzed with PMOD 3.8 software (PMOD Technologies LLC). Spherical volumes of interest (VOIs) of 2 mm radius were made within each sample to calculate the X-ray attenuation (in Hounsfield Units, HU) for each concentration. Average values of Milli-Q water and dispersions were used to calculate HU values in the images, with attenuation being 0 HU for water and –1000 HU for air.

The colloidal stability of nanoparticles suspensions in 50 mM MES solution at pH=6.5 (physiological pH simulator), having a nanoparticles content of 0.5 mg cm$^{-3}$, was monitored by analysing the particle size distribution (hydrodynamic diameter, $d_H$) obtained from dynamic light scattering (DLS) measurements (Zetasizer NanoZS90, Malvern).

Cell viability was analysed by the 3-(4,5-dimethylthiazol-2-yl)-2,5-diphenyltetrazolium bromide (MTT) colorimetric assay [28]. Vero cells (kidney epithelial cells from African green monkey) were acquired from the American Type Culture Collection (ATCC: CCL-81) and cultured at 37 °C in a 5% $CO_2$ atmosphere in Dulbecco's modified Eagle's medium (DMEM) supplemented with 10% foetal bovine serum (FBS), 2 mM glutamine and 100U cm$^{-3}$ penicillin/streptomycin.

For the MTT analysis, 5000 cells were seeded in each well of 96-well plates and grown for 24 h. After 24 h, the medium was replaced with fresh medium containing the nanoparticles in varying concentrations. After cultivation again for 24 h, 20 µL of MTT dye solution (5 mg cm$^{-3}$ in PBS) was added to each well. After 3 h of incubation at 37 °C and 5% $CO_2$, the medium was removed, the cells were washed with fresh medium, the plate was centrifuged at 2,500 rpm for 30 minutes, the supernatant was removed and formazan crystals were dissolved in 100 µL of DMSO. The absorbance of each well was read on a microplate reader (Biotek ELX800) at 570 nm. The relative cell viability (%) related to control wells containing cells without nanoparticles was calculated as [A]test/[A]control x100.

## 3. Results and Discussion

### 3.1. Undoped $NaLa(WO_4)_2$ nanoparticles

As it has been amply documented, the formation of uniform particles through precipitation requires a specific reaction kinetics that must be found by adjusting the experimental conditions such as reagents concentration, temperature, reaction time, nature of solvent and nature of additives [29]. In this work, we fixed the reagent concentrations (0.03 mol dm$^{-3}$ lanthanum (III) chloride and 0.16 mol dm$^{-3}$ sodium tungstate), temperature (120ºC) and aging time (20 h) and varied the nature of the solvent (EG/$H_2O$ mixtures with variable EG:$H_2O$ volumetric ratio) in order to find the effect of this parameter on the morphological characteristics of the precipitated particles. The results are summarized in Table 1 and Fig. 1a-c. As observed, in pure EG, irregular and aggregated nanoparticles with very small size were produced (Fig. 1a). However, when adding water to the EG, important differences in the morphological characteristics of the precipitates were detected. Thus, for an EG/H$_2$O volumetric ratio of 4/1, well-dispersed ellipsoidal particles with broad size distribution in the submicrometric range were obtained (Fig. 1b). A further increase of the amount of water added (EG/H$_2$O) = 3/2) gave rise to an increase of particle size up to the micron range, keeping the size heterogeneity (Fig. 1c). This behavior could be explained by the reduction of the viscosity (0.01733 Pa·s for ethylene glycol and 0.001 Pa·s, for water) and the rise of the dielectric constant (80 for water and 37 for EG at 20°C) of the mixed solvent as increasing the amount of water. Thus, the dropping of viscosity would favor the diffusion process of solutes, which consequently accelerates particles growth, explaining the increase of particle size observed as increasing the amount of water. Besides, the rise of the dielectric constant of the solvent must result in an increase of solubility of the precursors, lowering the supersaturation degree and, consequently, increasing the nuclei radius according to the Kelvin equation [30], which might also contribute to such particle size increase. Finally, the higher degree of dispersion of the precipitated particles detected as increasing the amount of water can be also attributed to the consequent increase of the dielectric constant, which must give rise to a higher zeta potential value and therefore, to a higher repulsion between the growing particles, disfavoring their agglomeration [31].

We have shown in previous papers that the addition of PAA to the starting solutions plays an important role in the formation of nanoparticles with controlled size and shape of different Ln-based compounds by homogeneous precipitation, since this polymer may act as a capping agent [32]. We adopted this strategy here in order to reduce particle size down to the nanometer range and improve the size uniformity. In effect, the addition of increasing amounts of PAA (up to 2 mg cm$^{-3}$) into the EG:H$_2$O solution (volumetric ratio 4/1) containing the tungstate precursor, gave rise to a progressive decrease of particle size (Table 1) and to a significant increase of size homogeneity (Fig. 1d,e). A further increase of the amount of PAA up to 4 mg cm$^{-3}$ did not have an appreciable effect on the size and morphology of the final nanoparticles (Fig. 1f and Table 1).

The role played by PAA on controlling the size of the precipitated particles is revealed by the FTIR spectrum (Fig. 2a, red line) of the ellipsoidal nanoparticles shown in Fig. 1e, which are chosen from now on as a representative example. The spectrum displays the absorptions associated to the W-O stretching vibration of the $WO_4^{2-}$ anion (<1000 cm$^{-1}$) [19], along with two broad bands at about 3400 and 1640 cm$^{-1}$, which are due to absorbed water. Some other weaker bands are detected in the 1710-1400 cm$^{-1}$ region, which are not present in the spectrum corresponding to the sample synthesized without PAA (Fig. 2a, blue line). Such features can be attributed to the C=O stretching (1711 cm$^{-1}$), $CH_2$ stretching (1450 cm$^{-1}$) and the symmetric (1400 cm$^{-1}$) and asymmetric (1545 cm$^{-1}$) stretching vibrations of the carboxylate ions of the PAA species, probably adsorbed on the particles surfaces [33]. The amount of such species was determined by TGA measurements (Fig. 2b). The TGA curve obtained for such functionalized sample showed a weight loss (4%) in the 25-300ºC range due to the release of adsorbed water and a further loss (3.7%) between 300 and 400ºC. The latter is absent in the TGA curve of a sample prepared in the absence of PAA (shown in Fig. 1b), being associated to the decomposition of PAA molecules.

The presence of PAA molecules on the surface of the nanoparticles could block the growth of the nanoparticles, explaining their ability to decrease the final size. The fact that the particle size does not vary by increasing the amount of PAA above 2 mg cm$^{-3}$ suggests that this amount is sufficient to achieve a complete PAA coating on the surface of the nanoparticles.

The ellipsoidal nanoparticles shown in Fig. 1e were identified by XRD as tetragonal NaLa(WO$_4$)$_2$ (PDF 1-79-1118, ICDD 2016) as illustrated in Fig. 2c. Crystallite size estimated using the Scherrer formula from the (200) reflection (2θ~33°) of the XRD patterns of this sample was found to be lower (25 nm) than the dimensions of the ellipsoids, which suggests that the nanoellipsoids were polycrystalline.

Therefore, the synthesis method developed in this work is suitable for the one-pot synthesis of uniform ellipsoidal NaLa(WO$_4$)$_2$ nanoparticles with tetragonal structure, and their simultaneous surface modification with PAA polymer.

### 3.2. Undoped NaLa(MoO$_4$)$_2$ nanoparticles

For the NaLa(MoO$_4$)$_2$ nanoparticles, we used the same synthesis conditions that yielded the NaLa(WO$_4$)$_2$ nanoellipsoids shown in Fig. 1e, but replacing the tungstate precursor for the molybdate one. Uniform particles with ellipsoidal shape were also obtained (Fig. 3a) in this case, although they presented a higher axial ratio (2.5) than those corresponding to the tungstate system (1.88) and a smaller size (115 x 45 nm). Such particles consisted of tetragonal NaLa(MoO$_4$)$_2$ (PDF 1-79-2243, ICDD 2016) (Fig. 3b). The crystallite size calculated using the Scherrer equation, from the width of the (200) reflection, was 16 nm indicating that the NaLa(MoO$_4$)$_2$ nanoparticles were also polycrystalline. FTIR spectroscopy confirmed the presence of PAA molecules on the particle surface (Fig. 3c), the amount of which was 4.8% by weight, as determined from the TGA curves (Fig. 3d). Such an amount is higher than that corresponding to the NaLa(WO$_4$)$_2$ system (3.7 %), which is in agreement with the lower surface area expected for the latter case, as a consequence of the higher particle size.

### 3.3. Eu$^{3+}$-doped NaLa(XO$_4$)$_2$ (X = W, Mo) nanoparticles

Eu$^{3+}$-doped NaLa(XO$_4$)$_2$ (X = W or Mo) nanoparticles were synthesized by using the experimental conditions involved in the synthesis of the undoped samples functionalized with PAA (NaLa(XO$_4$)$_2$@PAA) shown in Figs. 1e (NaLa(WO$_4$)$_2$) and 3a (NaLa(MoO$_4$)$_2$), but adding the desired amounts of the Eu$^{3+}$ precursor to the starting cations solutions. In these experiments, the total Ln concentration (0.03 mol dm$^{-3}$) was kept constant and the doping level was varied (from 3 to 27% molar) in order to optimize the luminescent properties of these nanophosphors.

It was found that the doping procedure did not produce important changes in the morphological characteristics of the precipitated nanoparticles whose shape was similar (Figs. S1 and S2) to that obtained for the undoped systems. However, a progressive decrease of particle size as increasing the doping level was detected (Table 2). This effect has also been noticed for other Ln doped nanoparticles with different composition [34], and may be associated with variations in the precipitation kinetics and/or the unit cell volume induced by the doping cations [35].

The success of the doping process was manifested by the composition (Eu/(Eu+La) mol ratio) of the samples measured by ICP, which in all cases, was very close to the nominal values (Table 2). EDX mapping analyses revealed that the Eu$^{3+}$ cations were homogeneously distributed across the nanoparticles as shown in Figs. 4 and S3 for the Eu:NaLa(WO$_4$)$_2$ and Eu:NaLa(MoO$_4$)$_2$ systems, respectively.

The XRD patterns of the doped samples (Figs. S4 and S5) were also similar to that of undoped systems, indicating that they also crystallized into the tetragonal phase. No significant variations in crystallite size were either observed for the doped samples respect to those of the undoped ones (Table 3) confirming the polycrystalline character for all samples.

The unit cell parameters obtained for the doped samples (Table 3) confirmed the incorporation of the Eu cations to the tungstate and molybdate matrices forming a solid solution. Thus, a progressive contraction of the unit cell was detected as increasing the amount of the doping cation (Fig. 5) in agreement with the smaller radius in eightfold coordination of $Eu^{3+}$ (1.066 Å) when compared with that of $La^{3+}$ (1.16 Å) [36].

Finally, the FTIR spectra of all doped samples (Fig. S6) also showed weak bands in the 1710-1400 cm$^{-1}$ region, confirming that PAA molecules also covered the doped nanoparticles.

*3.4. Luminescence Properties*

The excitation spectrum recorded for an aqueous suspension of the Eu(7%):NaLa(WO$_4$)$_2$@PAA nanoparticles, monitored at the most intense $Eu^{3+}$ emission band (610 nm), is presented in Fig. 6. The spectrum displays a strong and broad excitation band, with maximum at 262 nm. This band correspond to the reported an energy transfer (ET) process from the excited levels of $WO_4^{2-}$ group to the $Eu^{3+}$ cations [24]. The spectrum shows, in addition, much weaker bands resulting from the f-f electronic transitions characteristic of the $Eu^{3+}$ ions [24], the most intense appearing at 393 nm due to the $^7F_0$-$^5L_6$ transition (inset in Fig. 6).

The excitation spectrum of an aqueous suspension of the 7% Eu:NaLa(MoO$_4$)$_2$@PAA nanoparticles (Fig. 6) was very similar to that of the tungstate based nanophosphor, except that the band associated to the energy transfer from the molybdate groups to the $Eu^{3+}$ cations appeared at lower energy (290 nm), in agreement with the literature [37], and showed higher intensity. The latter finding seems to manifest that the molybdate matrix absorbs energy and/or transfers it to the $Eu^{3+}$ cations in a more effective way that the tungstate one.

Upon excitation of the Eu:NaLa(WO$_4$)$_2$@PAA nanoparticles through the ET band ($\lambda_{ex}$=262 nm), the emission bands characteristic of the electronic transitions from the $^5D_0$ to the $^7F_J$ (J = 1, 2, 3, and 4) levels of $Eu^{3+}$ were detected (Fig. 7a). These bands were very similar to those detected for the Eu:NaLa(MoO$_4$)$_2$@PAA system (Fig. 7b), as expected, since the electronic transitions of $Ln^{3+}$ cations are hardly affected by crystal field. It should be noted that the relative intensities of the $^5D_0$-$^7F_2$ (610 nm) and $^5D_0$-$^7F_1$ (590 nm) transitions strongly depend on the local symmetry of the $Eu^{3+}$ ions. Thus, the latter transition is magnetic-dipole allowed and insensitive to the local environment of the $Eu^{3+}$ cations, whereas the $^5D_0$-$^7F_2$ electric-dipole transition becomes the strongest one when the Eu local symmetry is lowered [38]. In our samples, the emission originated from the $^5D_0$-$^7F_2$ transitions is clearly dominant irrespective of the matrix, indicating that the $Eu^{3+}$ occupies a crystallographic site in the host lattice with no inversion center as that of $La^{3+}$ cations in the scheelite structure with $S_4$ symmetry [39]. Such emission is responsible for the strong red luminescence of the samples observed in the photographs inserted in Fig. 7, which correspond to the samples containing 20 % Eu.

The intensity of the emissions for each system is plotted in Fig. 7c, as a function of the Eu content. It can be observed that in both cases, the increase of the $Eu^{3+}$ doping level from 3% to 20%, gave rise to a progressive rise of the intensity of the emission as a consequence of the larger number of emission centers. However, for higher doping levels (27%), the intensity of luminescence remained almost unaltered. This behavior suggests the presence of the well-known concentration quenching effect [40], which was further studied by evaluating the luminescent dynamics of the samples.

The decay curves obtained for the $^5D_0 \rightarrow {}^7F_2$ $Eu^{3+}$ transition (610 nm), after pulsed excitation at the ET band, as function of Eu concentration for both systems are presented in Figs. 8a and 8b. In all cases, the curves could be fitted by using a bi-exponential temporal dependence according to Equation (1):

$$I(t) = I_{01} \exp(-t/\tau_1) + I_{02} \exp(-t/\tau_2) \qquad (1)$$

where $I(t)$ is the luminescence intensity, $t$ is the time after excitation and $\tau_i$ $(i = 1, 2)$ is the decay time of the $i$-component, with initial intensity $I_{0i}$. The corresponding fitting parameters are presented in Table 4. This bi-exponential behavior is common in Ln-based nanophosphors. The shorter lifetime values are usually associated to the transitions of Ln cations located on the nanoparticles outer-layers, and therefore, influenced by quenching processes due to interactions with solvent molecules and impurities, whereas the longer lifetime values are due to transitions of the Ln cations located in the bulk, and therefore, less affected by such quenching processes. Table 4 also shows that the relative weight of the long component increases with increasing doping level in both systems, suggesting that the relative amount of $Eu^{3+}$ cations located in the bulk also increases.

The average lifetime values, $\langle\tau\rangle$, calculated as

$$\langle\tau\rangle = \frac{\int_{t_0}^{t_f} tI(t)dt}{\int_{t_0}^{t_f} I(t)dt} = (\tau_1^2 I_1 + \tau_2^2 I_2)/(\tau_1 I_1 + \tau_2 I_2)$$

(2)

where $t_f$ represents the time required for the luminescence signal to reach the background, are also included in Table 4. It is important to mention that these values (from 1106 to 959 µs, for the $Eu:NaLa(WO_4)_2$ samples; and from 861 to 655 µs, for the $Eu:NaLa(MoO_4)_2$ samples) are much higher than those previously reported for these systems (from 401 to 640 µs, for $Eu:NaLa(WO_4)_2$ [19, 21, 26, 41,] and from 400 to 620 µs, for $Eu:NaLa(MoO_4)_2$ [42,43]). This finding seems to indicate the higher efficiency of our nanophosphors.

The $\langle\tau\rangle$ values have also been plotted in Fig. 8c as a function of the $Eu^{3+}$ content. For both systems, $\langle\tau\rangle$ kept decreasing during the whole range of studied compositions, confirming the concentration quenching effect in such range. It should be noted that the increase of the emissions intensity detected when increasing the Eu content up to 20% can be explained by considering that the increase of luminescent centers compensates, at least in part, the concentration quenching effect. Therefore, from the practical point of view, the samples having a 20% Eu can be considered as the optimum ones in view of their highest emission intensity for which, they were used for further analyses.

Finally, the observation of Fig. 7c also reveals that the intensity of the emissions is clearly higher for the Eu:NaLa(MoO$_4$)$_2$ nanophosphors when compared with those based on Eu:NaLa(WO$_4$)$_2$, disregarding the Eu content, which is in agreement with the higher intensity of the energy transfer band observed for the first system (Fig. 6). To explain this behavior, the luminescence quantum yield (QY) of the nanophosphors is also plotted in Fig. 7c as a function of the Eu content. Specifically, we measured the so called *overall quantum yield* by exciting the samples through their corresponding ET bands, which therefore contains information on the intrinsic quantum yield as well as on the efficacy with which energy is transferred from the matrix to the Eu$^{3+}$ cations [44]. First, it was observed that the QY values vs. Eu content followed a similar trend that the luminescent intensity for both systems, confirming that the optimum samples were those with the highest doping level (20 %). Interestingly, it was also found that the QY values were higher for the molybdate-based system than those for the tungstate one, irrespective of the Eu content. This behavior seems to indicate that the higher intensity of the Eu:NaLa(MoO$_4$)$_2$ nanophosphors may be due to either a higher intrinsic QY of the luminescence of Eu$^{3+}$ in the molybdate matrix and/or to a higher efficiency of ET in this matrix. These results allow concluding that, because of their better luminescence properties, the molybdate-based nanophosphors here developed are better candidates for their use as luminescent probes for bioimaging.

*3.5. X-ray Attenuation Properties*

The X-ray attenuation values of aqueous suspensions containing different concentrations of the optimum nanophosphors (Eu(20%):NaLa(MoO$_4$)$_2$ and Eu(20%):NaLa(WO$_4$)$_2$) were measured and compared with aqueous suspensions containing the same concentrations of Iohexol, which is a clinically approved CT contrast agent. The corresponding CT phantom images are shown in Fig. 9a, where a higher contrast can be observed for the nanophosphors suspensions than for Iohexol at any concentration. The X-ray attenuation values, in Hounsfield units (HU), obtained from these images are plotted in Fig. 9b versus the nanophosphors concentration. As observed, a good linear correlation between the HU values and the concentration of the suspensions was obtained for all systems. In addition, it was observed that the slope of the line was higher for the Eu:NaLa(WO$_4$)$_2$ sample (37.1 HU/mg cm$^{-3}$) than for the Eu:NaLa(MoO$_4$)$_2$ one (24.2 HU/mg cm$^{-3}$), and it was, in both cases, significantly higher than that of Iohexol (15.0 HU/mg cm$^{-3}$). These results are as expected since the X-ray attenuation coefficient increases as increasing the average atomic number of the elements constituting the sample [45]. This behavior confirms a superior CT imaging performance for our nanophosphors when compared with Iohexol, especially for the Eu:NaLa(WO$_4$)$_2$ case, which can be considered as promising CT contrast agents.

*3.6. Colloidal Stability and Biocompatibility*

The colloidal stability of the optimum nanophosphors (Eu(20%)-doped NaLa(WO$_4$)$_2$@PAA and Eu(20%)-doped NaLa(MoO$_4$)$_2$@PAA), was evaluated to assess their suitability for biomedical applications. It was observed that the hydrodynamic diameter (d$_h$) obtained from DLS measurements (Fig. 10) in freshly prepared nanoparticles suspensions in MES buffer at physiological pH (MES 50 mM at pH=6.5), was only slightly higher (160 nm, for the tungstate system and 140 nm for the molybdate system) than the mean particle size obtained from the TEM pictures (100 x 60 nm, for tungstate and 105 x 31 nm for molybdate). This behavior clearly indicates that the nanoparticles can be well dispersed in MES buffer without significant aggregation.

Biocompatibility studies for both optimum nanophosphors were performed using Vero cells through the MTT assay [28]. The metabolic activity of the cells was measured after 24 hours culture and the results are shown in Fig. 11. As observed, irrespective of the matrix, the nanoparticles showed negligible toxicity effects with viability percentages as high as >90% for concentrations up to 0.3 mg cm$^{-3}$.

Therefore, in the tested conditions, our nanoparticles meet the colloidal stability and biocompatibility criteria required for their use in biomedical applications.

## 4. Conclusions

We have developed a procedure for the synthesis, for the first time, of uniform and hydrophilic Eu$^{3+}$-doped sodium lanthanum tungstate nanoparticles with ellipsoidal shape, based on the aging at 120ºC for 20 h of solutions containing, in an ethylene glycol/H$_2$O mixed solvent, specific amounts of lanthanum chloride, europium chloride, sodium tungstate and polyacrylic acid. This polymer is essential to control particle size and shape and remains anchored to the nanoparticles surface, which becomes functionalized with carboxylate groups. This procedure is also suitable (obviously changing the tungstate by the molybdate precursor) for the synthesis of isostructural sodium lanthanum molybdate nanoparticles with similar morphological characteristics. Both kinds of phosphors emit intense red light when excited with UV radiation through the energy transfer band from the tungstate or molybdate anions to the Eu$^{3+}$ cations. In both cases, the samples showing the most intense emission were those doped with 20 mol % Eu$^{3+}$, for which they were considered as the optimum luminescent probes. Nevertheless, the luminescence intensity of the molybdate-based samples was higher than that of the tungstate-based ones for any of the compositions assayed. This behaviour, probably due to either a higher intrinsic quantum yield of the $Eu^{3+}$ luminescence in the molybdate matrix and/or to a higher efficiency of energy transfer in this matrix, manifests the superior performance of the molybdate system from the luminescent point of view. The X-ray absorption capacity of the tungstate-based material was higher than that of the molybdate-based one, in both cases being notably higher than that of a commercial probe (Iohexol), indicating that they are better contrast agents for X-ray computed tomography, especially, the tungstate system. Finally, the cell viability of both type of probes was very high and their colloidal stability at physiological pH was acceptable. Because of these properties, the developed nanomaterials may find potential uses as bifunctional probes for luminescent bioimaging and X-ray computed tomography.

## Acknowledgements

This work has been supported by the Spanish Ministry of Science, Innovation and Universities (RTI2018-094426-B-I00), the European Research Council (ERC) under the European Union's Horizon 2020 research and innovation program (NANOPHOM, grant agreement no. 715832), DGA and Fondos Feder (Bionanosurf E15_17R) and CSIC (PIC2016FR1). This work was also supported in part by Siemens Healthcare S.L.U. MM thanks MINECO for Juan de la Cierva Fellowship. We also acknowledge the use of the CNA's ICTS NanoCT facilities and Dr. F.M. Varela-Feria (CITIUS – Universidad de Sevilla) for EDX experiments.

**Table 1.** Shape and size of the particles obtained by aging at 120ºC for 20 h, EG/H$_2$O (different ratios) solutions containing LaCl$_3$ (0.03 mol dm$^{-3}$), Na$_2$WO$_4$ (0.16 mol dm$^{-3}$) and different amounts of PAA. Standard deviations are included in parentheses. Corresponding TEM micrographs are also indicated in the last column.

| EG/H$_2$O v/v | [PAA] mg cm$^{-3}$ | Shape | Length (nm) | Width (nm) | TEM |
|---|---|---|---|---|---|
| 5/0 | 0 | Irregular | | | Fig.1a |
| 4/1 | 0 | Ellipsoidal | heterogeneous | heterogeneous | Fig. 1b |
| 3/2 | 0 | Ellipsoidal | heterogeneous | heterogeneous | Fig. 1c |
| 4/1 | 1 | Ellipsoidal | 190 (60) | 95 (25) | Fig. 1d |
| 4/1 | 2 | Ellipsoidal | 168 (20) | 89 (11) | Fig. 1e |
| 4/1 | 4 | Ellipsoidal | 162 (19) | 78 (10) | Fig. 1f |

**Table 2.** Nominal and experimental (measured by ICP) Eu/(Eu+La) molar ratios and size (from TEM) for the Eu:NaLa(XO$_4$)$_2$@PAA (X = W, Mo) nanoparticles. Standard deviation is included in parentheses.

| System | Eu/(Eu+La) (% nominal) | Eu/(Eu+La) (ICP) | Length (nm) | Width (nm) |
|---|---|---|---|---|
| Eu:NaLa(WO$_4$)$_2$ | 0 | - | 168 (20) | 89 (11) |
| | 3 | 3.0 | 150 (19) | 80 (10) |
| | 7 | 7.4 | 120 (16) | 70 (8) |
| | 14 | 14.6 | 100 (14) | 60 (6) |
| | 20 | 20.8 | 80 (11) | 55 (5) |
| | 27 | 28.1 | 70 (10) | 40 (5) |
| Eu:NaLa(MoO$_4$)$_2$ | 0 | - | 115 (14) | 45 (6) |
| | 3 | 2.9 | 106 (12) | 32 (5) |
| | 7 | 7.2 | 102 (10) | 34 (5) |
| | 14 | 14.6 | 105 (11) | 31 (6) |
| | 20 | 21.0 | 80 (9) | 30 (5) |
| | 27 | 27.9 | 75 (8) | 25 (5) |

**Table 3.** Unit cell parameters, cell volume and crystallite size determined for the NaLa(XO$_4$)$_2$@PAA (X= W, Mo) samples doped with different amounts of Eu$^{3+}$ (nominal values). Experimental errors are included in parentheses.

| System | Eu/(Eu+La) (% nom.) | a=b (Å) | c (Å) | Cell volume (Å$^3$) | Crystallite size (nm) |
|---|---|---|---|---|---|
| Eu:NaLa(WO$_4$)$_2$ | 0 | 5.3548(3) | 11.671(1) | 334.6452 | 25 |
| | 3 | 5.3526(3) | 11.662(1) | 334.1105 | 26 |
| | 7 | 5.3494(3) | 11.651(1) | 333.4028 | 28 |
| | 14 | 5.3425(3) | 11.631(1) | 331.9948 | 27 |
| | 20 | 5.3364(3) | 11.612(1) | 330.6808 | 25 |
| | 27 | 5.3301(3) | 11.591(1) | 329.2728 | 26 |
| Eu:NaLa(MoO$_4$)$_2$ | 0 | 5.3447(4) | 11.737(1) | 335.2886 | 16 |
| | 3 | 5.3421(3) | 11.732(1) | 334.8068 | 16 |
| | 7 | 5.3384(7) | 11.721(1) | 334.0567 | 16 |
| | 14 | 5.3321(3) | 11.701(1) | 332.6703 | 17 |
| | 20 | 5.3260(3) | 11.683(1) | 331.4139 | 16 |
| | 27 | 5.3198(3) | 11.662(1) | 330.0708 | 16 |

**Table 4.** Lifetime values (τ$_1$ y τ$_2$) and amplitudes (A$_1$ y A$_2$) corresponding to the two components and average lifetime (<τ>) determined, for NaLa(WO$_4$)$_2$@PAA and NaLa(MoO$_4$)$_2$@PAA nanoparticles doped with different amounts of Eu$^{3+}$, from the decay curves registered for the $^5D_0 - {}^7F_2$ transition (610 nm). Standard errors in the calculation of the average lifetimes are below 1%.

| System | Eu/(Eu+La) (% nominal) | A$_1$ | τ$_1$ (μs) | A$_2$ | τ$_2$ (μs) | <τ> (μs) |
|---|---|---|---|---|---|---|
| Eu:NaLa(WO$_4$)$_2$ | 3 | 43 | 731 | 57 | 1268 | 1106 |
| | 7 | 27 | 457 | 73 | 1158 | 1069 |
| | 14 | - | - | 100 | 1017 | 1035 |
| | 20 | - | - | 100 | 1003 | 1005 |
| | 27 | - | - | 100 | 959 | 959 |
| Eu:NaLa(MoO$_4$)$_2$ | 3 | 40 | 482 | 60 | 985 | 861 |
| | 7 | 32 | 426 | 68 | 893 | 807 |
| | 14 | 22 | 333 | 78 | 792 | 744 |
| | 20 | 19 | 315 | 81 | 744 | 705 |
| | 27 | 13 | 201 | 87 | 675 | 655 |

**Captions**

**Fig. 1.** TEM images of the particles obtained by aging at 120°C for 20 h solutions containing $LaCl_3$ (0.03 mol $dm^{-3}$) and $Na_2WO_4$ (0.16 mol $dm^{-3}$) using different solvents: a) pure EG, b) EG/$H_2O$ (4:1) and c) EG/$H_2O$ (3:2); and using an EG/$H_2O$ mixture (4:1) as solvent in the presence of different amounts of PAA: d) 1 mg $cm^{-3}$, e) 2 mg $cm^{-3}$ and f) 4 mg $cm^{-3}$

**Fig. 2.** a) FTIR spectra of the nanoellipsoids obtained in presence (Fig. 1b) and in the absence (Fig. 1e) of PAA; b) TGA curves of the nanoparticles obtained in the presence (Fig. 1b) and in the absence (Fig. 1e) of PAA; c) X-ray diffraction pattern of the nanoparticles shown in Fig, 1e; the PDF corresponding to the $NaLa(WO_4)_2$ tetragonal phase is also included.

**Fig. 3.** TEM micrograph (a), XRD pattern (b), FTIR spectrum (c) and TGA curve (d) of the particles obtained by aging at 120°C for 20 h solutions containing $LaCl_3$ (0.03 mol $dm^{-3}$), $Na_2MoO_4$ (0.16 mol $dm^{-3}$) and PAA (2 mg $cm^{-3}$) in a EG/$H_2O$ (4:1) mixture. The PDF corresponding to the $NaLa(MoO_4)_2$ tetragonal phase is also included in 3b.

**Fig. 4.** EDX mapping obtained for a single particle of the Eu (20%):$NaLa(WO_4)_2$ sample.

**Fig. 5.** Evolution of unit cell volume with Eu content for the $Eu^{3+}$:$NaLa(WO_4)_2$@PAA (red) and $Eu^{3+}$:$NaLa(MoO_4)_2$@PAA (blue) nanoparticles.

**Fig. 6.** Excitation spectra of $Eu^{3+}$(7%)-doped $NaLa(WO_4)_2$@PAA (red line) and $Eu^{3+}$ (7%)-doped $NaLa(MoO_4)_2$@PAA (blue line) nanoparticles. The inset is a magnification of the 350-425 nm spectral region.

**Fig. 7.** Emission spectra of the of the $Eu^{3+}$:$NaLa(WO_4)_2$@PAA (a) and $Eu^{3+}$:$NaLa(MoO_4)_2$@PAA (b) nanoparticles with different Eu contents. The insets correspond to photographs of the Eu(20%)-doped $NaLa(WO_4)_2$@PAA (a) and Eu(20%)-doped $NaLa(MoO_4)_2$@PAA (b) nanoparticles suspensions taken under UV (254 nm) illumination, c) Integrated intensity of the emissions (balls) and overall quantum yield

(stars) of the Eu$^{3+}$:NaLa(WO$_4$)$_2$@PAA (red) and Eu$^{3+}$:NaLa(MoO$_4$)$_2$@PAA (blue) nanophosphors as a function of the Eu$^{3+}$ doping level. Lines are guides for the eye.

**Fig. 8.** Temporal decays for the $^5D_0 \rightarrow {}^7F_2$ transition (610 nm) of the Eu-doped NaLa(WO$_4$)$_2$@PAA (a) and NaLa(MoO$_4$)$_2$@PAA (b) nanoparticles with different Eu content and average lifetime values calculated from such decays (c).

**Fig. 9.** CT phantom images (a) and HU values (b) for the Eu(20%):NaLa(WO$_4$)$_2$@PAA and Eu(20%):NaLa(MoO$_4$)$_2$@PAA samples and for Iohexol at different concentrations in water.

**Fig. 10.** Hydrodynamic diameter ($d_h$) obtained from DLS measurements for the Eu(20%):NaLa(MoO$_4$)$_2$@PAA and Eu(20%):NaLa(MoO$_4$)$_2$@PAA nanoparticles dispersed in 50 mM MES solution at pH = 6.5.

**Fig. 11.** Cell vialibility of the Eu(20%):NaLa(MoO$_4$)$_2$@PAA and Eu(20%):NaLa(MoO$_4$)$_2$@PAA nanoparticles.

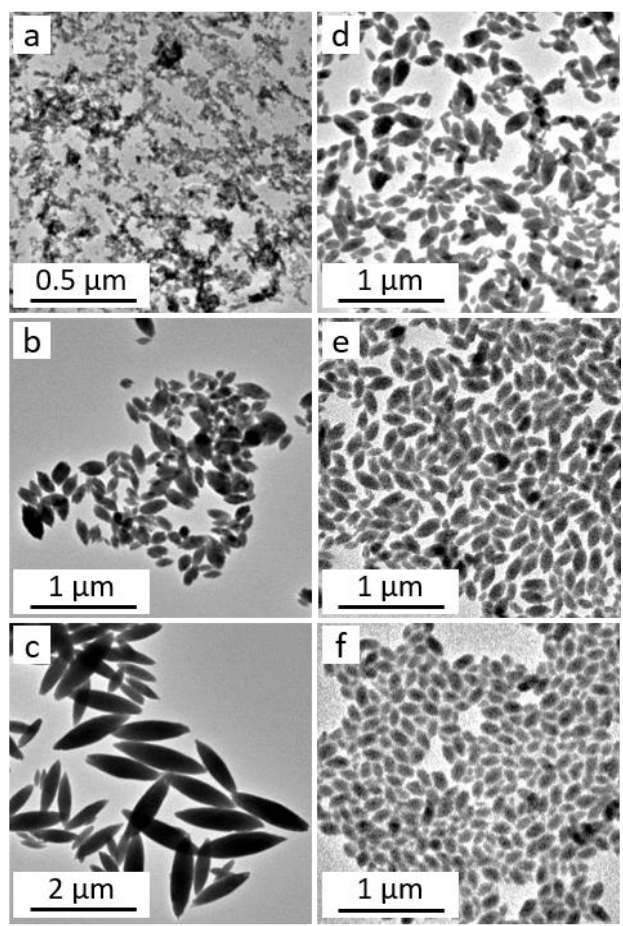

Fig. 1

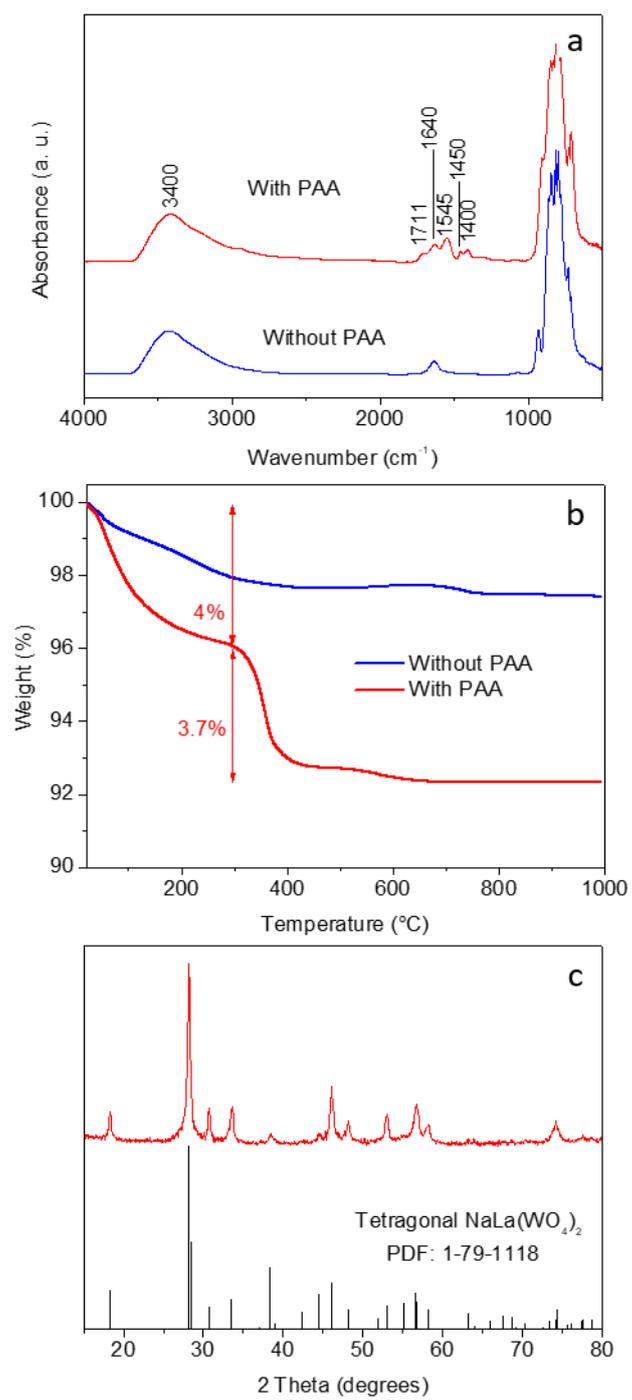

Fig. 2

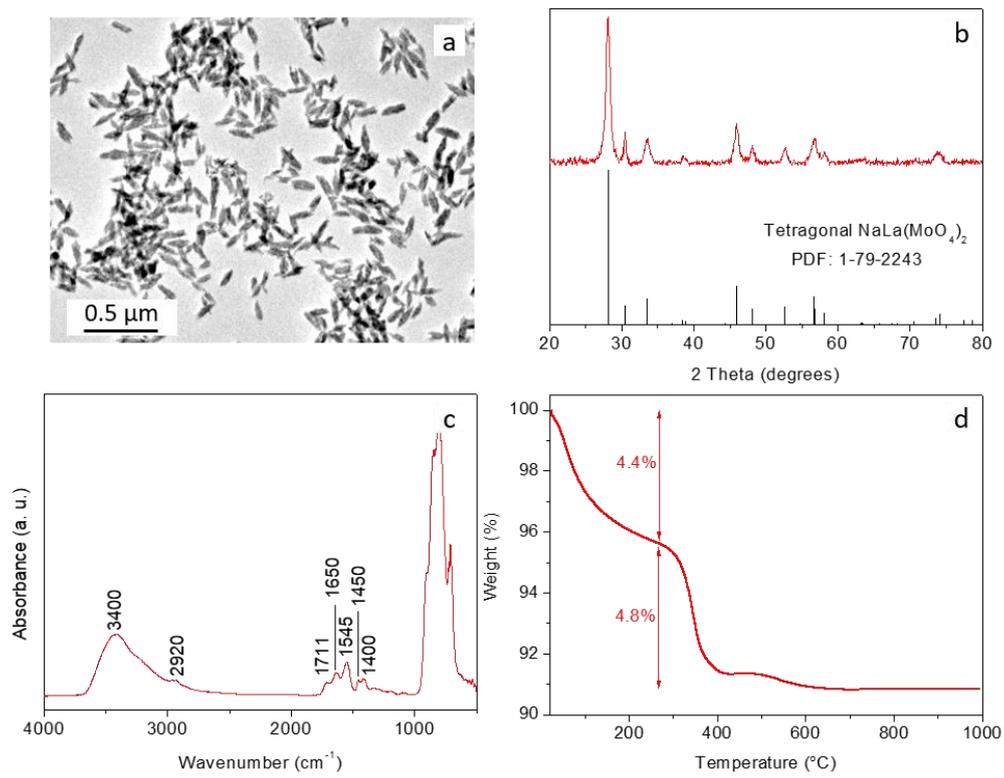

Fig. 3

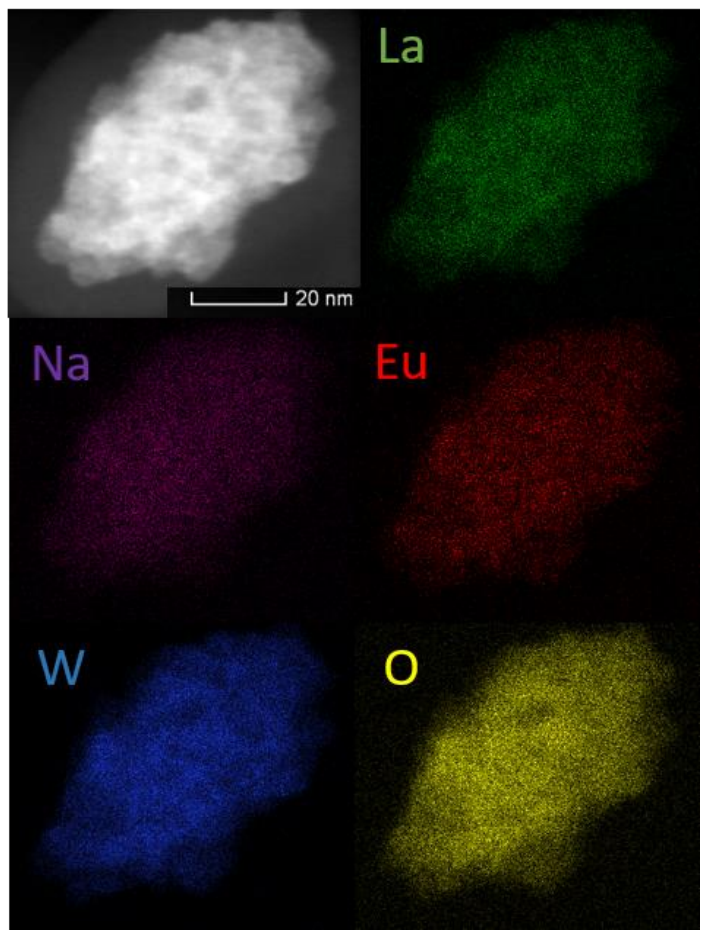

Fig. 4

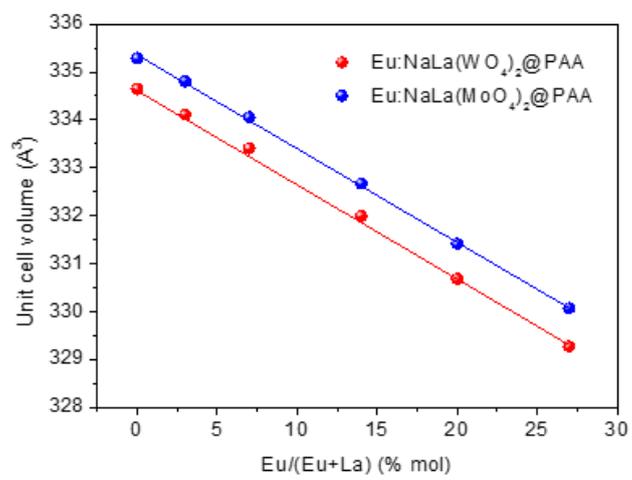

Fig. 5

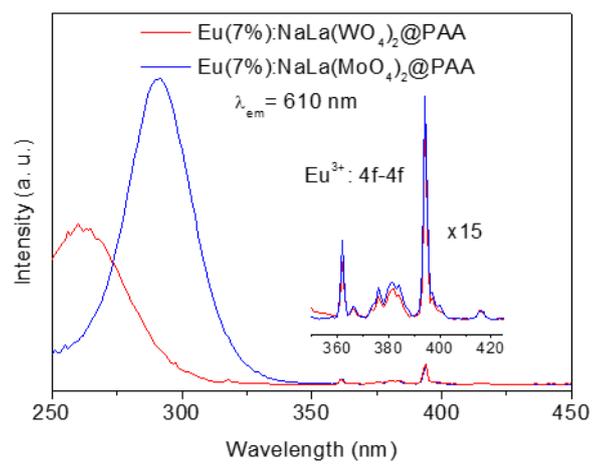

Fig. 6

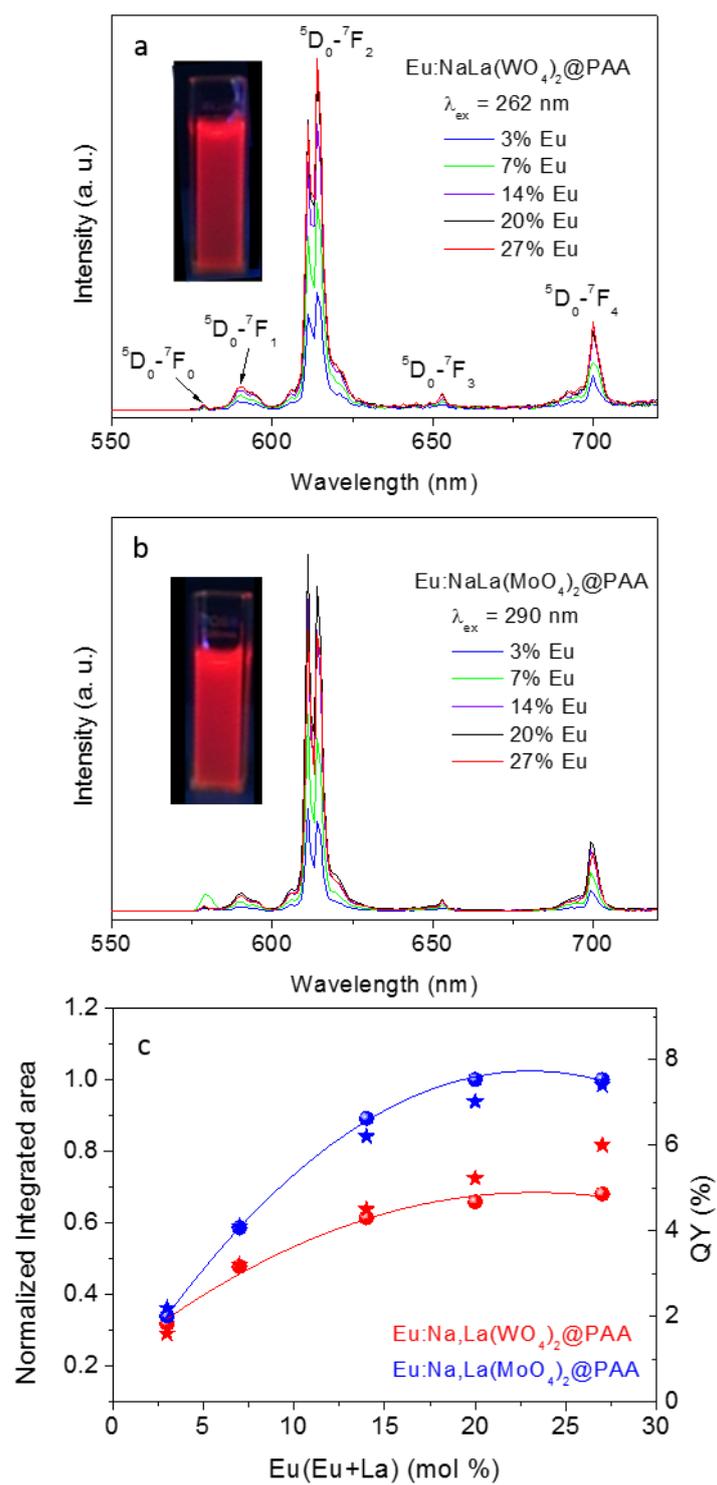

Fig. 7

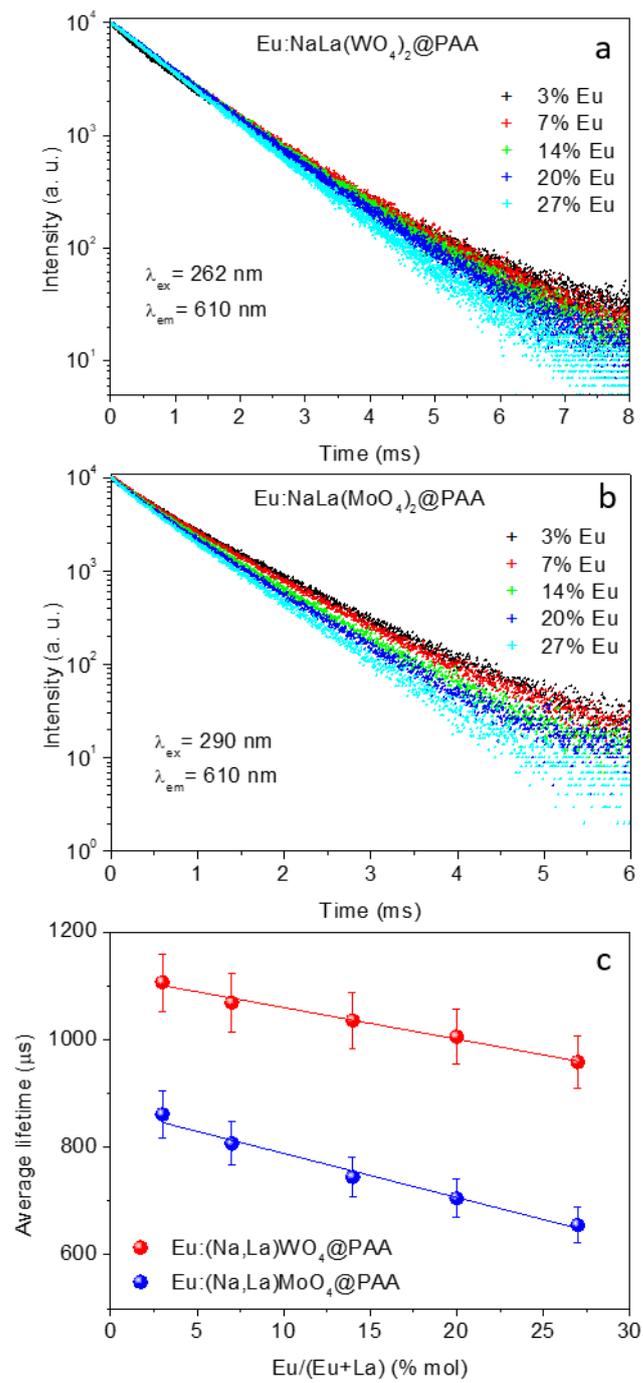

Fig. 8

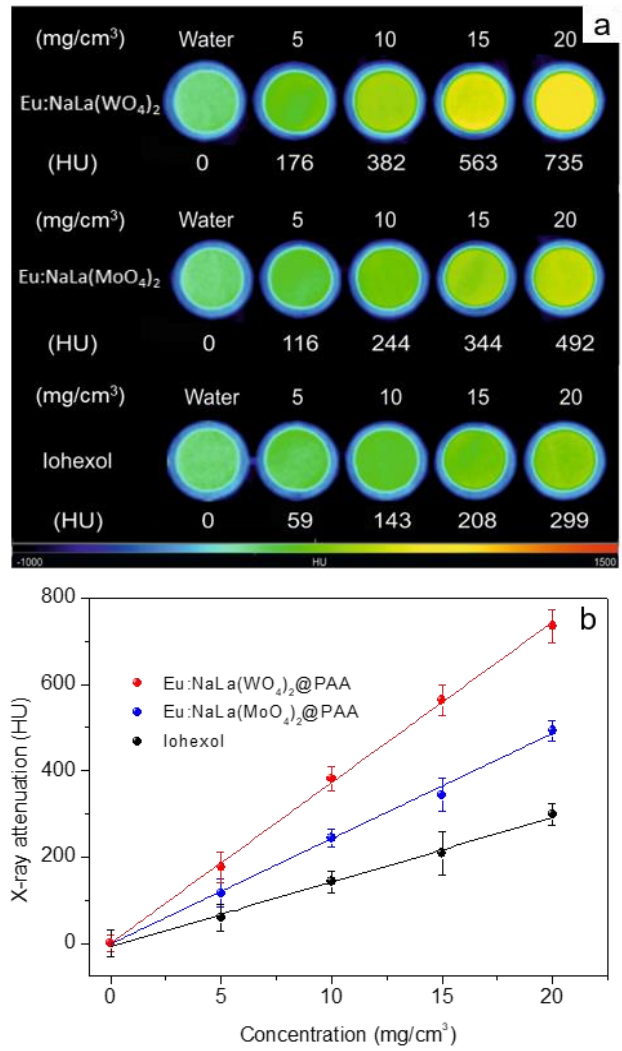

Fig. 9

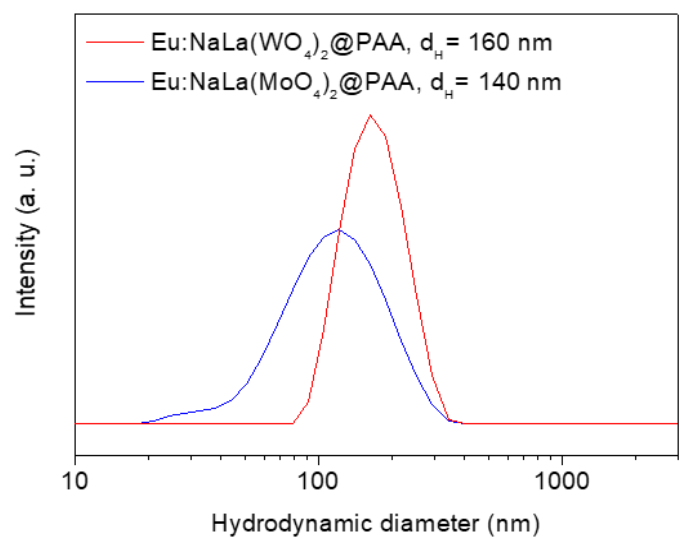

Fig. 10

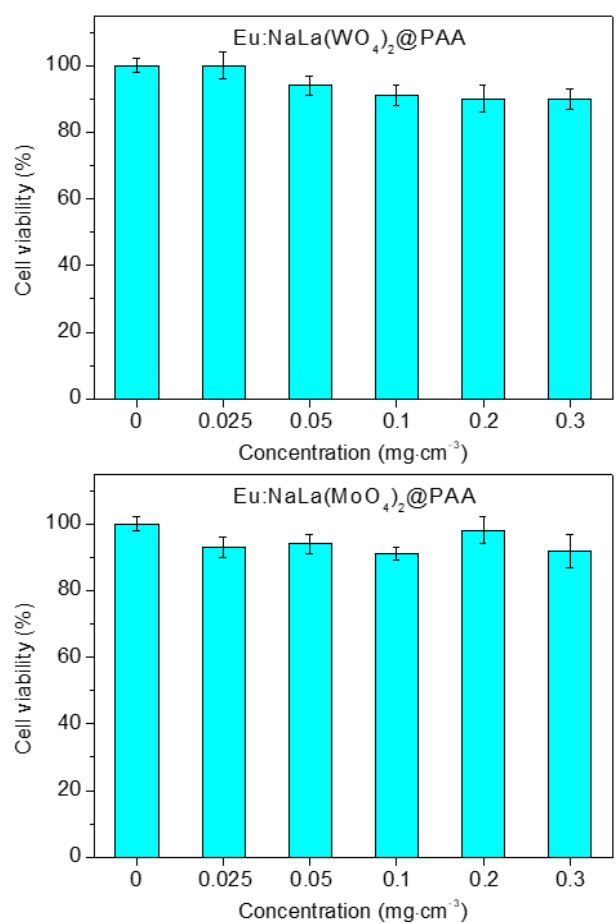

Fig. 11